\documentclass[12pt]{iopart}
\usepackage{graphicx}

\begin{document}

\title[The Richardson effect and a transient solution of pipe flow]{The Richardson's Annular effect and a transient solution of oscillating pressure-driven flow in circular pipes}

\author{F J Camacho$ˆ1$, R E Martinez$ˆ1$, L Rendon$ˆ2$}

\address{$ˆ1$Departamento de Fisica, Universidad Nacional de Colombia, Bogota}

\address{$ˆ2$ Departamento de Matematicas, Universidad Nacional de Colombia, Bogota}

\ead{fjcamachor@unal.edu.co}

\begin{abstract}
In this paper it is shown that the location of the characteristic overshoot of the Richardson's annular effect changes with the Kinematic Reynolds number  $\omega^*=\omega r_0^2 /\nu$ in the range of frequencies within the laminar regime. From the study of the Richardson's overshoot at different times it was identified the existence of aparent transverse damped waves similar to those ones observed in the famous Stokes second problem, the physical analysis of this waves was used for the establishment of a semi-empirical law that gives the functional relation of the mean overshoot maximum with the kinematic reynolds number, say $B_0(\omega^*)=2.28 + 0.51{\omega^*}^{-1/2}$.  
Finally a transient solution was constructed and verified asymptotically for large times, and  the tipical time for which the transient solution resembles the steady oscillating one was identified to be dependent of the viscosity of the fluid and of the radius of the pipe.
\end{abstract}


\maketitle

\section{Introduction}
Nowadays there is a little bunch of analytical solutions on transient flows, tipically this kind of solutions  doesn't satisfy the initial condition of rest flow at t=0, including those ones related to pipe flows, specifically consider the well known velocity field of circular pipe flow due to an oscillating pressure gradient.

\begin{equation}
\frac{d P}{dz}=-\rho K e^{i \omega t}.
\end{equation}

Is clear that circular pipe flow motivates the use of cylindrical coordinates for the Navier Stokes equations. Now two fundamental hypotheses will be taken, first a regime in which the angular component of velocity vanishes $u_{\phi}=0$, and second a flow that doesn't change in the axial component, say $z$; this means that the remaining components of velocity are given with the following dependences:

\begin{equation}
\eqalign{
u_r=u_r(r,t) \cr
u_z=u_z(r,t)}  
\end{equation}

The continuity equation in cylindrical coordinates under the two previous hypotheses gives:

\begin{equation}
\frac{1}{r}\frac{\partial}{\partial r} (\rho r u_r)=0
\end{equation}

Where $\rho$ is the fluid constant density. Now the no-slip condition for the radial component together with (3) show that the radial velocity must vanish everywhere inside the pipe, giving a one-directional flow  given by the axial component $u_z(r,t)$. Now the Navier Stokes equations in cylindrical coordinates reduces to the simplified axial equation:

\begin{equation}
 \rho\left(\frac{\partial u_z}{\partial t}+u_r\frac{\partial u_z}{\partial r}+u_z\frac{\partial u_z}{\partial z}\right)= 
   -\frac{\partial P}{\partial z} +\mu\left(\frac{1}{r}\frac{\partial}{\partial r}\left(r\frac{\partial u_z}{\partial r}\right)+\frac{\partial^2 u_z}{\partial z^2}\right). 
\end{equation}

All the derivatives in $z$ give zero using the second assumption of axial independence, and the remaining non-linear term in (4) vanish due to the fact that the continuity equation (3) gives $u_r=0$; finally writting $u_z:=u$ the axial momentum equation is:

\begin{equation}
\rho \frac{\partial u}{\partial t}=-\frac{dP}{dz} +\mu\left(\frac{\partial^2 u}{\partial r^2}+\frac{1}{r} \frac{\partial u}{\partial r}\right).
\end{equation}

The most natural way to solve the steady oscillating state of (5), for which the flow is already oscillating with the same frequency of the external pressure gradient, is by introducing an harmonic velocity field:

\begin{equation}
u(r,t)=R(r) e^{i \omega t}.
\end{equation}
 
 By using the no-slip condition at pipe wall $u(r_0,t)=0$, where $r_0$ is the pipe radius, is easy to check that the variable separation assumed above leads to the following velocity field \cite{W}:

\begin{equation}
u(r,t)=\frac{K}{i \omega} e^{i \omega t} \left(1-\frac{J_0(r \sqrt{-i \omega / \nu})}{J_0(r_0 \sqrt{-i \omega/ \nu})}\right).
\end{equation}

 However this flow doesn't satisfy the initial condition of rest state at $t=0$ just as was mentioned before; evidently only in the case when the pressure gradient starts with a value equals to zero the initial condition is satisfied, this is equivalent to have a sinusoidal gradient like $sin(\omega t)$, but it is desirable to have a solution that starts from rest for a general harmonic pressure gradient as in (1).

\section{Richardson's Annular effect}

The particular oscillating flow described by (7) exhibits a quite interesting behavior known as the Richardson annular effect \cite{W}, which consists of the presence of a velocity overshoot near to the wall, this can be observed for high frequencies of the pressure gradient in pipe flow experiments; Sexl \cite{S} observed theoretically this overshoot by expanding the solution (7) for large frecuencies, thereby he achieved the following asymptotic fluid flow:

\begin{equation}
\frac{u(r^*,t)}{u_{max}}=\frac{4}{\omega^*}\left[sin(\omega t)-\frac{e^{-B}}{\sqrt{r^*}}sin(\omega t -B)\right].
\end{equation}

Where $\omega^*=\omega r_0^2 /\nu$ is the kinematic Reynolds number of the problem, $r^*=r/r_0$ is a dimensionless radial coordinate, $u_{max}=K r_0 / 4 \nu$ is the centerline velocity  for steady Poiseuille flow with pressure gradient -$\rho K$, and B is defined in terms of $r_0$ and $\omega^*$ as follows:

\begin{equation}
B=(1-r^*)\sqrt{\frac{\omega^*}{2}}.
\end{equation}

\begin{figure}[htp]
	\centering 
		\includegraphics[scale=0.5]{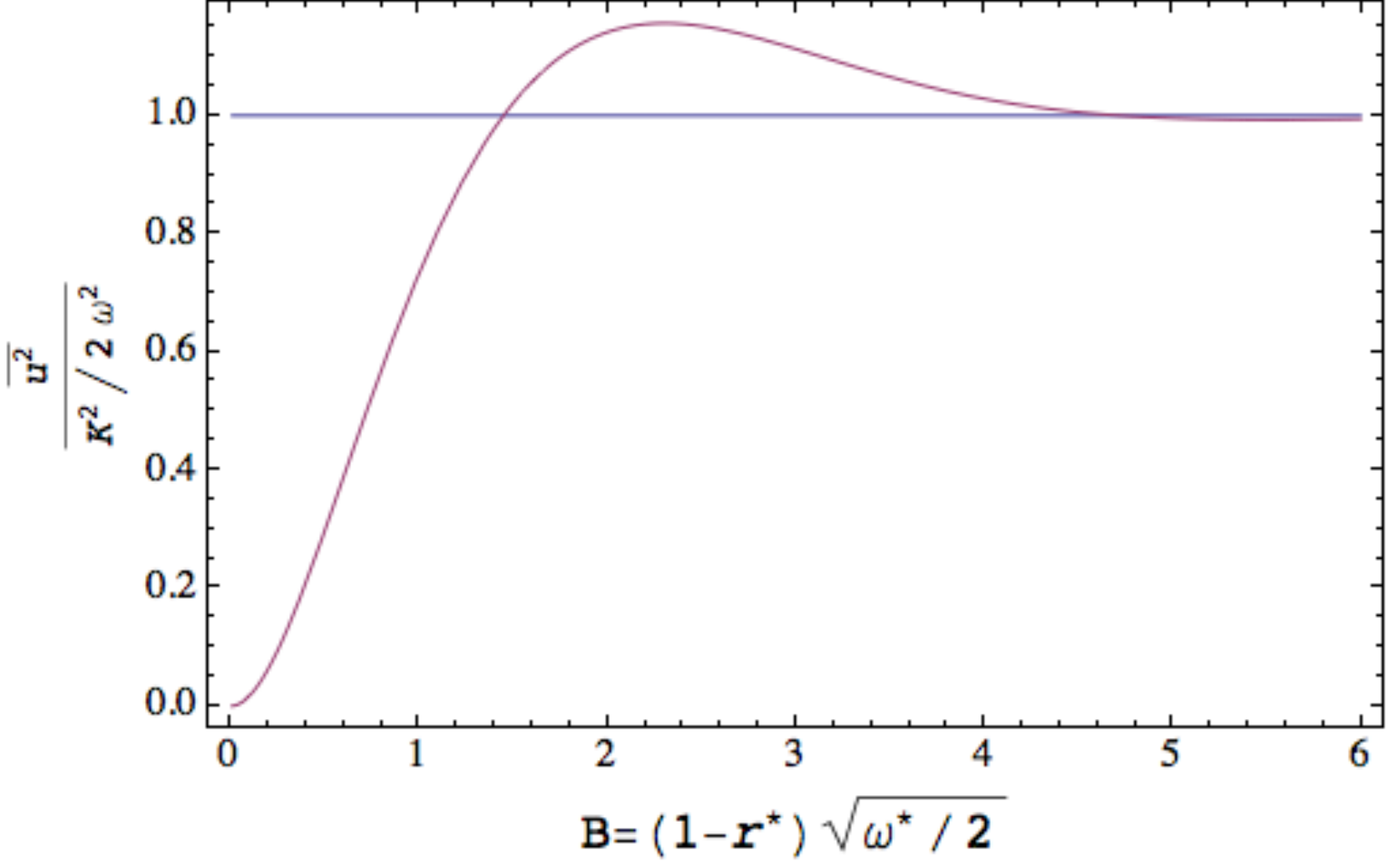}
\caption{ Plot of the Near Wall velocity overshoot for the specific case $\omega^*=500$ }
	\label{fig:un}
\end{figure}

A plot of the mean square velocity derived from the asymptotic flow (8) is given in figure 1 from which is easy to see the overshoot  for B near to zero, that means that the overshoot occurs near to the wall of the pipe. 

The mean squared velocity was used instead of the mean velocity because due to the oscillating character of the last one the mean velocity gives  $\overline{u}=0$.

\section{Transverse Damped waves in oscillating flow}

A plot of six velocity profiles given by (8) and equally time spaced over a period of oscillation for a kinematic Reynolds number $\omega^*=500$ is given in figure 2.

\begin{figure}[htp]
	\centering 
		\includegraphics[scale=0.5]{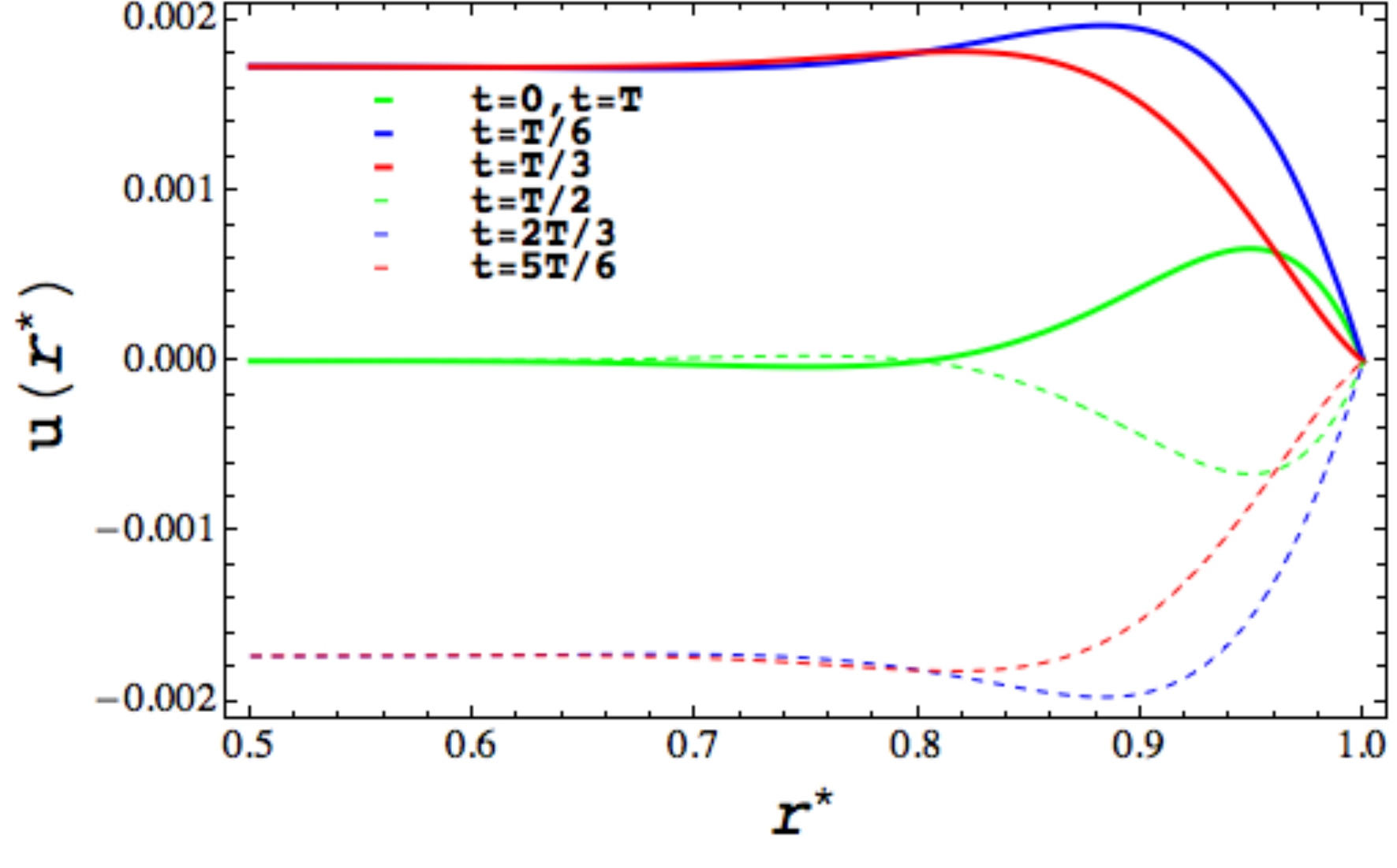}
\caption{Equally time spaced velocity profiles over a period of oscillation}
	\label{fig:un}
\end{figure}

It clearlly shows that the location of the instantaneous maximum of the overshoot changes with time. To observe with detail the location of maximum, the derivative of the velocity field in (8) was calculated:

\begin{equation}
u'(r^*,t):=\frac{\partial u(r^*,t)}{\partial r^*}=\frac{-4 u_{max}}{\omega^*}\frac{\partial}{\partial r^*}\left(\frac{e^{-B}}{\sqrt{r^*}}sin(\omega t -B)\right).
\end{equation}

Here the vanishing point of the derivative given by $u'(r^*_{max},t)=0$, indicates the location of the maximum. Plots of this function for the same times used in figure 2 for the velocity profiles are shown in figure 3 \footnote{The crossings with the horizontal axis correspond to the solutions of $u'(r^*,t)=0$}. They suggest that the overshoot maximum moves somehow periodically with time, being the period of the motion the half-period of the pressure gradient $\pi/\omega$. 

During a half period, as time increases the value of $r_{max}^*$ decreases, that means the the maximum of the overshoot is moving from the wall to the centerline of the pipe; then after the half-period the maximum of the overshoot appears again near to the wall of the pipe.

It seems that it returns to its original position after a half-period of the pressure gradient (see intersection of continous and dashed lines of the same color in figure 3).

\begin{figure}[htp]
	\centering 
		\includegraphics[scale=0.5]{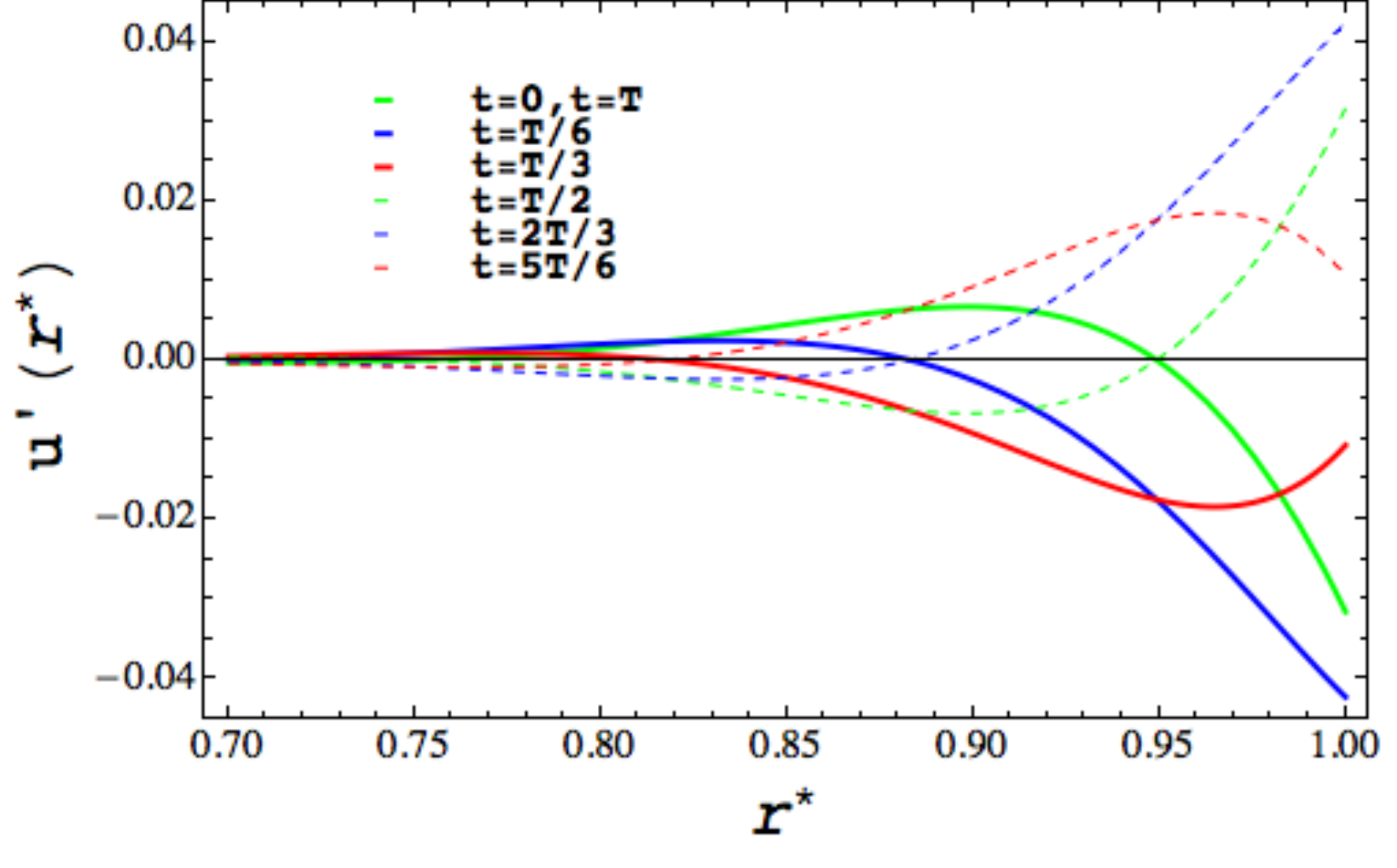}
\caption{Equally time spaced plots of $u'(r^*)$ over a semi-period of oscillation.}
	\label{fig:un}
\end{figure}

Indeed the location of the overshoot maximum $r^*_{max}$ is quite dificult to define in this case since the equation that define it as a function of time is given by a trully complicated non linear relation that arises from the condition $u'(r^*_{max},t)=0$ which turns out to be:

\begin{equation}
tan(\omega t -B)= \frac{e^{-B}}{\sqrt{r^*}}\frac{dB}{d r^*}\left(\frac{d}{d r^*}(e^{-B}/\sqrt{r^*})\right)^{-1}.
\end{equation}

This equation has multiple inverse functions because for a fixed time there exist infinite locations where $u'(r^*,t)$ vanishes; it can be seen that each location and therefore each root of (11) moves along the $r^*$ axis towards zero as times passes, eventually one of this roots may take a value smaller than $r^*_0=1$ which is the limit for this roots to have physical significance. Then what is really happening is that as time move forward an initial overshoot moves radially from the boundary towards the centerline of the pipe; as long as this overshoot moves it decays and become negligible, at the same time another new overshoot appears in the wall of the pipe and follow the same history of the previous one and so on. This analysis suggest a motion that behaves like some kind of wave  moving radially to the centerline of the pipe.  

Actually this behavior corresponds to transverse decaying waves in the pipe. The word "waves" in this case is quite dangerous because this a problem of difussion instead of a wave propagation one, so these are not real waves in the physical sense, they just show that the diffusion effects inside the pipe propagate in a oscillating way\footnote{Here it was followed the interpretation of Kundu \& Cohen \cite{Kun}, where they discuss the oscillating plate flow that resembles the same kind of wave appearance.}, however in what follows, for the sake of simplicity, we will use the term wave to refer to this phenomenon. 

The existence of these waves was noted by Stokes for the so-called oscillating plate flow, where he proposed a depth of penetration for them as \cite{Sch}:

\begin{equation}
\delta=\sqrt{\frac{2\nu}{\omega}}.
\end{equation}

Hence the amplitude of these waves is damped by a exponential factor $e^{-r/\delta}$,  this can be seen for this case of pipe flow from the velocity field (8) where we see a exponential contribution $e^{-(1-r^*)\sqrt{\omega^*/2}}$ ,which gives always positive values of $(1-r^*)$ inside the pipe, this can be identified with the radial distance traveled by the waves from the wall and this leads for our case to the same length defined by Stokes. 

Now it can be analized the behavior of the mean location of the maximum overshoot by computing the derivative of the mean squared velocity $\overline{u^2}$; so the mean location can be founded from:

\begin{equation}
\frac{d}{d r^*}\left(1-\frac{2e^{-B}}{\sqrt{r^*}}cos(B)+\frac{e^{-2B}}{r^*}\right)=0.
\end{equation}

Let $B_0(\omega^*)$ be the solution of this equation as a function of the Kinetic Reynolds number; this equation is non-linear so an analytical solution for the unknown $B_0(\omega^*)$ cannot be given. 

Now the depth of penetration defined in (12) changes like $O(\omega^{-1/2})$ and it is plausible to assume that the mean location of the overshoot given by (13) is directly related with the depth of penetration, so in a tentative way it can be proposed a functional form for $B_0$ as follows:

\begin{equation}
B_0(\omega^*)=A+\frac{B}{\sqrt{\omega^*}}.
\end{equation}

Numerical solutions of (13) for different frequencies were obtained (see dots in figure 4), also a non-linear least squares analisis using the function (14) was applied to the numerical solutions of (13), giving the parameters $A=2.28$ and $B=0.51$, the fitted function (dashed line) and the computed data matches quite good as shown in figure 4, the correlation coeficient of the fitted curve gives $R^2=0.99998$.

\begin{figure}[htp]
	\centering 
		\includegraphics[scale=0.5]{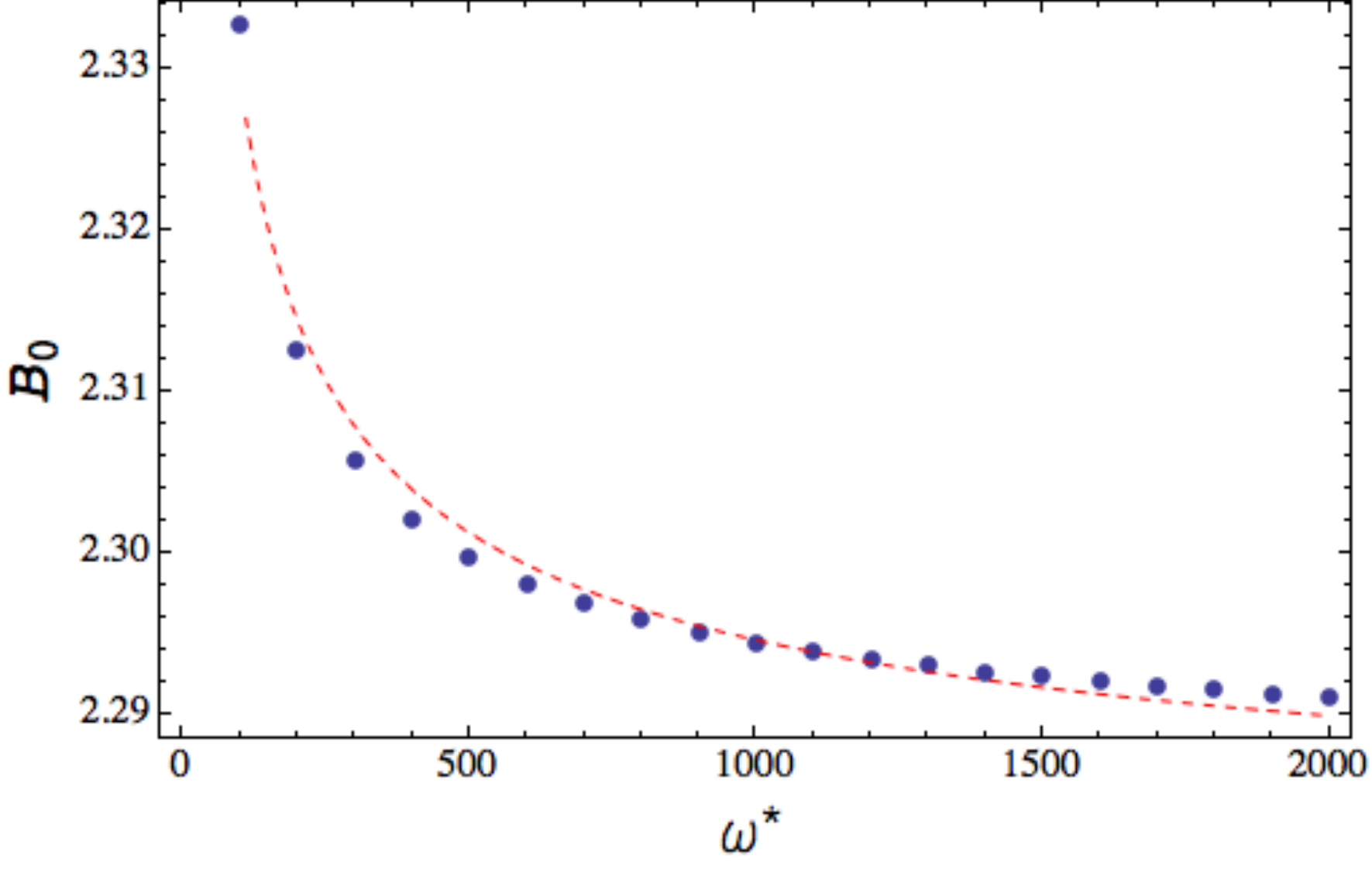}
\caption{Mean overshoot maximum as a function of the kinetic Reynolds number, the dots represent the numerical solutions of (13) for different values of $\omega^*$ and the dashed line is the function (14) adjusted by the non-linear regression.}
	\label{fig:un}
\end{figure}

\section{Unsteady decaying State}

To obtain a solution of (5) that satisfies the initial condition of fluid flow at rest when t=0, is necessary to assume a more general ansatz, motivated by the form of the homogeneous solution given when $d P/ dx=0$ \cite{Eq}, thereby it is proposed a solution of the form:

\begin{equation}
u(r^*,t)=\sum_{n=1}^{\infty} a_n(t) J_0(j_n r^*)
\end{equation}

Where $J_0(x)$ is a Bessel function of order zero of the first kind and $j_n$ is the nth zero of  $J_0(x)$. Now introducing (15) in (5) gives:

\begin{equation}
\sum_{n=1}^{\infty}\left(a_n'(t)+\frac{\nu j_n^2}{r_0^2}a_n(t)\right)J_0(j_n r^*)=K e^{i\omega t},
\end{equation}

This turns out to be the Fourier-Bessel expansion of $Ke^{i \omega t}$, hence the coeficients of the expansion are given by \cite{Eq2}:

\begin{equation}
a_n'(t)+\frac{\nu j_n^2}{r_0^2}a_n(t)=K e^{i \omega t} \frac{\int_0^1 xJ_0(j_n x) xd}{\int_0^1 x J_0^2(j_n x) dx}.
\end{equation}

Solving the integrals above by using some well known identities gives:

\begin{equation}
a_n'(t)+\frac{\nu j_n^2}{r_0^2}a_n(t)=\frac{2K e^{i\omega t}}{j_n J_1(j_n)}.
\end{equation}

The last relation is a first order ordinary differential equation fo the unknown coeficients of the expansion $a_n(t)$; now the initial condition $u(r^*,0)=0$ implies that $a_n(0)=0$ for all $n$ in the sum, therefore the solution of this initial value problem is quite easy and gives:

\begin{equation}
a_n(t)=2K\frac{e^{i \omega t}-e^{-\nu j_n^2t/r_0^2}}{(i\omega+\nu j_n^2/r_0^2)j_n J_1(j_n)}.
\end{equation}

Finally the desired solution is given by:

\begin{equation}
u(r^*,t)=2K\sum_{n=1}^{\infty}\frac{e^{i \omega t}-e^{-\nu j_n^2t/r_0^2}}{(i\omega+\nu j_n^2/r_0^2)j_n J_1(j_n)}J_0(j_n r^*).
\end{equation}

This solution gives rest flow at $t=0$, and for t $\rightarrow$ $\infty$ the real exponentials vanishes and we obtain a steady oscillating state:

 \begin{equation}
u(r^*,t)\sim 2Ke^{i \omega t}\sum_{n=1}^{\infty}\frac{J_0(j_n r^*)}{(i\omega+\nu j_n^2/r_0^2)j_n J_1(j_n)}.
\end{equation}

Rigorously, in order to have the steady oscillating solution (7) we should have the equality:

 \begin{equation}
\sum_{n=1}^{\infty}\frac{J_0(j_n r^*)}{(i\omega+\frac{\nu j_n^2}{r_0^2})j_n J_1(j_n)}=\frac{1}{2 i \omega}\left(1-\frac{J_0(r \sqrt{-i \omega / \nu})}{J_0(r_0 \sqrt{-i \omega/ \nu})}\right).
\end{equation}

Here we give numerical proof that indeed both expresions agree for all values of $r^*$,  where it were defined the following functions:

\begin{equation}
S_{\omega}^N(r^*)=\sum_{n=1}^{N}\frac{J_0(j_n r^*)}{(i\omega+\frac{\nu j_n^2}{r_0^2})j_n J_1(j_n)},
\end{equation}

\begin{equation}
D_{\omega}^N(r^*)=S_{\omega}^N(r^*)-\frac{1}{2 i \omega}\left( 1-\frac{J_0(r \sqrt{-i \omega / \nu})}{J_0(r_0 \sqrt{-i \omega/ \nu})} \right)
\end{equation}

Therefore the equality (22) is equivalent to the following limits:

\begin{equation}
\lim_{N \to \infty} \Re{\left(D_{\omega}^N(r^*)\right)}=0
\end{equation}

\begin{equation}
\lim_{N \to \infty} \Im\left(D_{\omega}^N(r^*)\right)=0
\end{equation}

Where this limit must hold $\forall r^*\in [0,1]$. Numerical evidence that support the convergence of these limits is shown in figures 5 and 6 for a frequency $\omega=100$, where it can be seen that as N increases the real and imaginary parts of $D_\omega^N(r^*)$ approach to zero.

Now notice that (20) have a superposition of decaying exponentials each one with a diferent tipical time of decaying depending on the corresponding $j_n$, so we have a decreasing sucesion of decaying times:

\begin{equation}
\tau_n=\frac{r_0^2}{\nu j_n^2}.
\end{equation}

All decaying contributions vanish faster than the lowest one of them given by $\tau_1$, so one could consider that a tipical time for which we recover the steady oscillating state solution must be $\tau_1$.

\begin{figure}[htp]
	\centering 
		\includegraphics[scale=0.5]{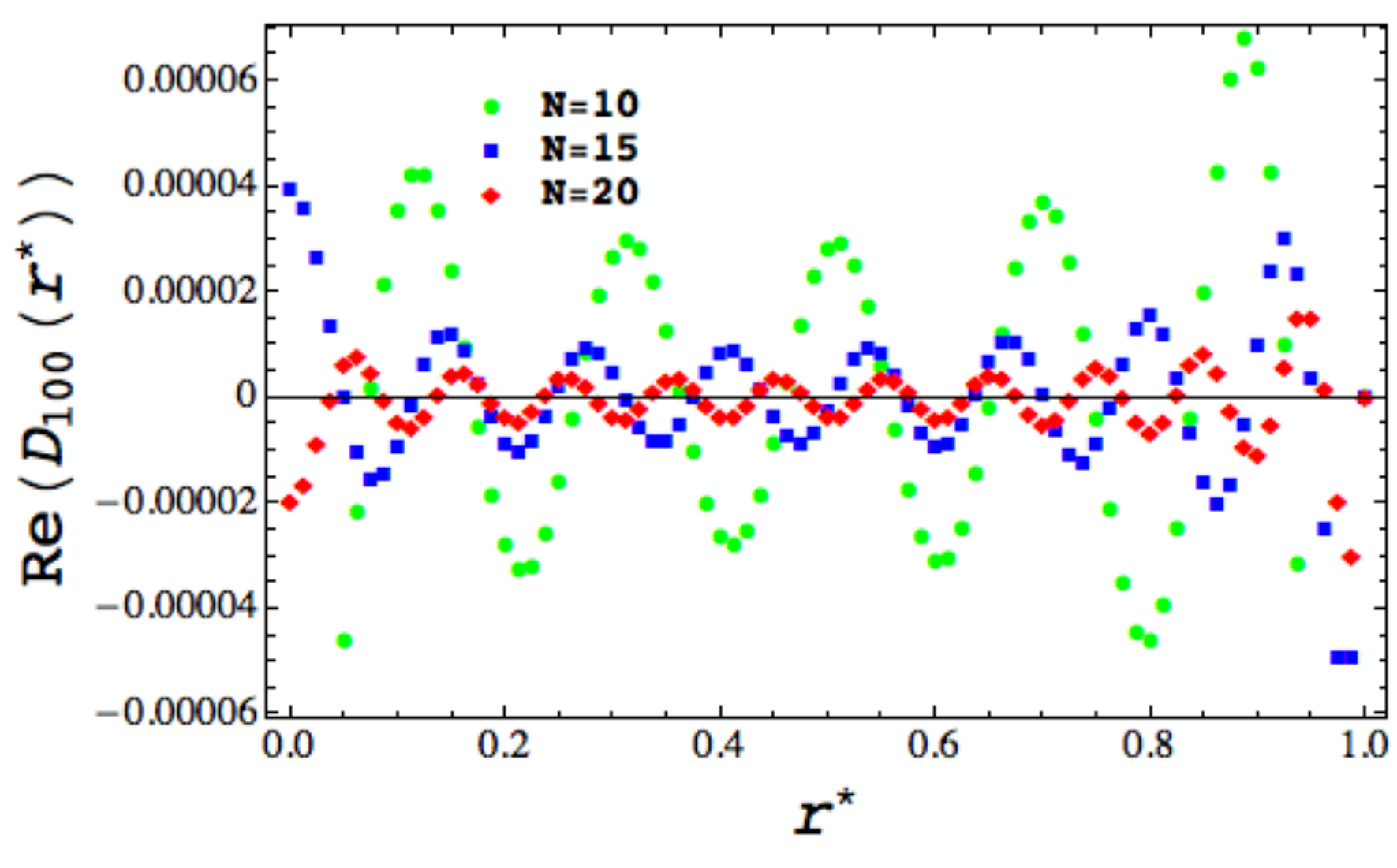}
\caption{Convergence of the Real part of the Difference funcion $D_{\omega} (r^*)$ as $N$ increases for $\omega=100$.}
	\label{fig:un}
\end{figure}

\begin{figure}[htp]
	\centering 
		\includegraphics[scale=0.5]{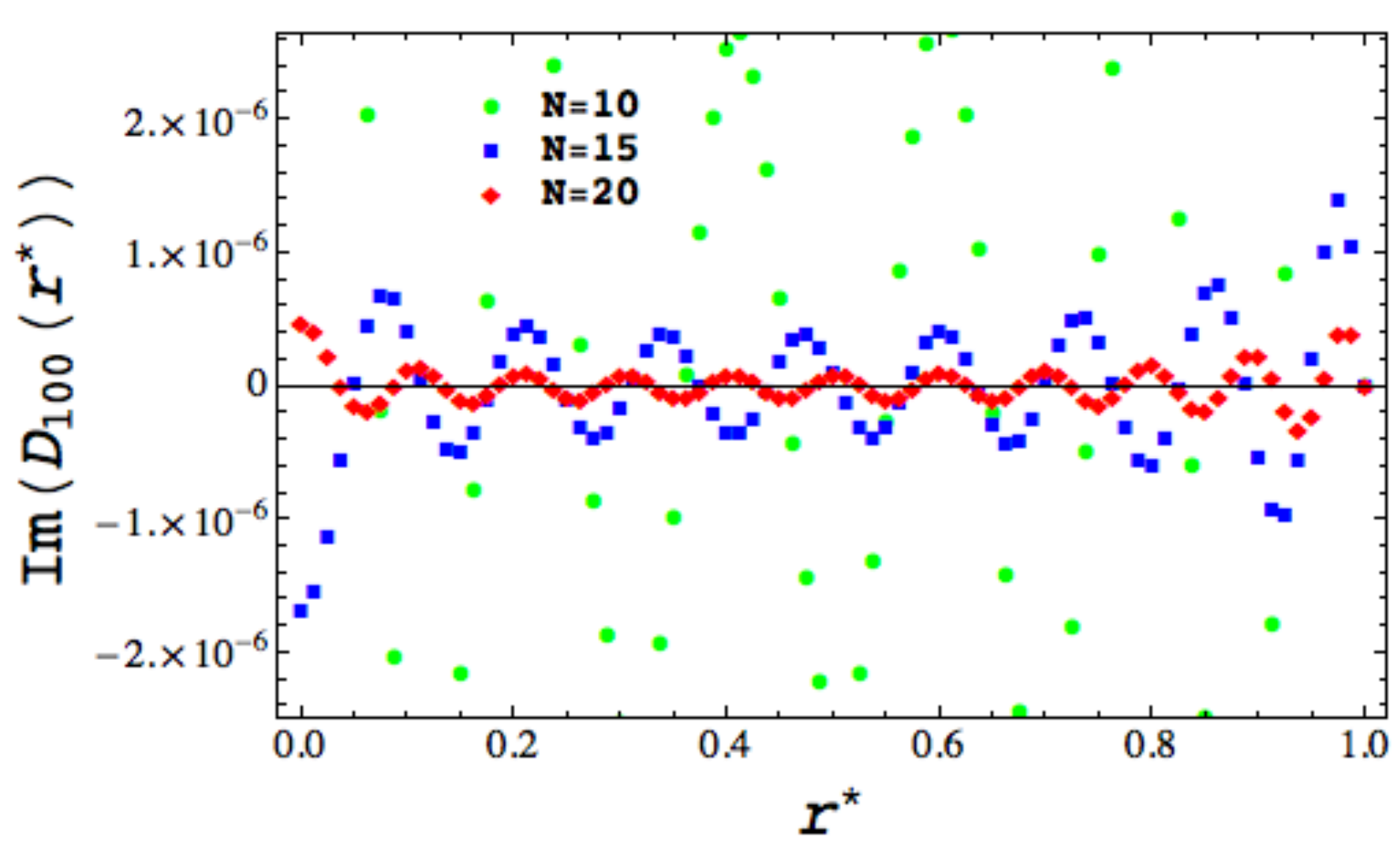}
\caption{Convergence of the Imaginary part of the Difference funcion $D_{\omega} (r^*)$ as $N$ increases for $\omega=100$.}
	\label{fig:un}
\end{figure}

In Figure 7 can be seen the comparison between the transient and the steady oscillating solutions for a time $t=\tau_1$, showing the agreement of both solutions at the tipical time for a frequency of $\omega=100$.

\begin{figure}[htp]
	\centering 
		\includegraphics[scale=0.5]{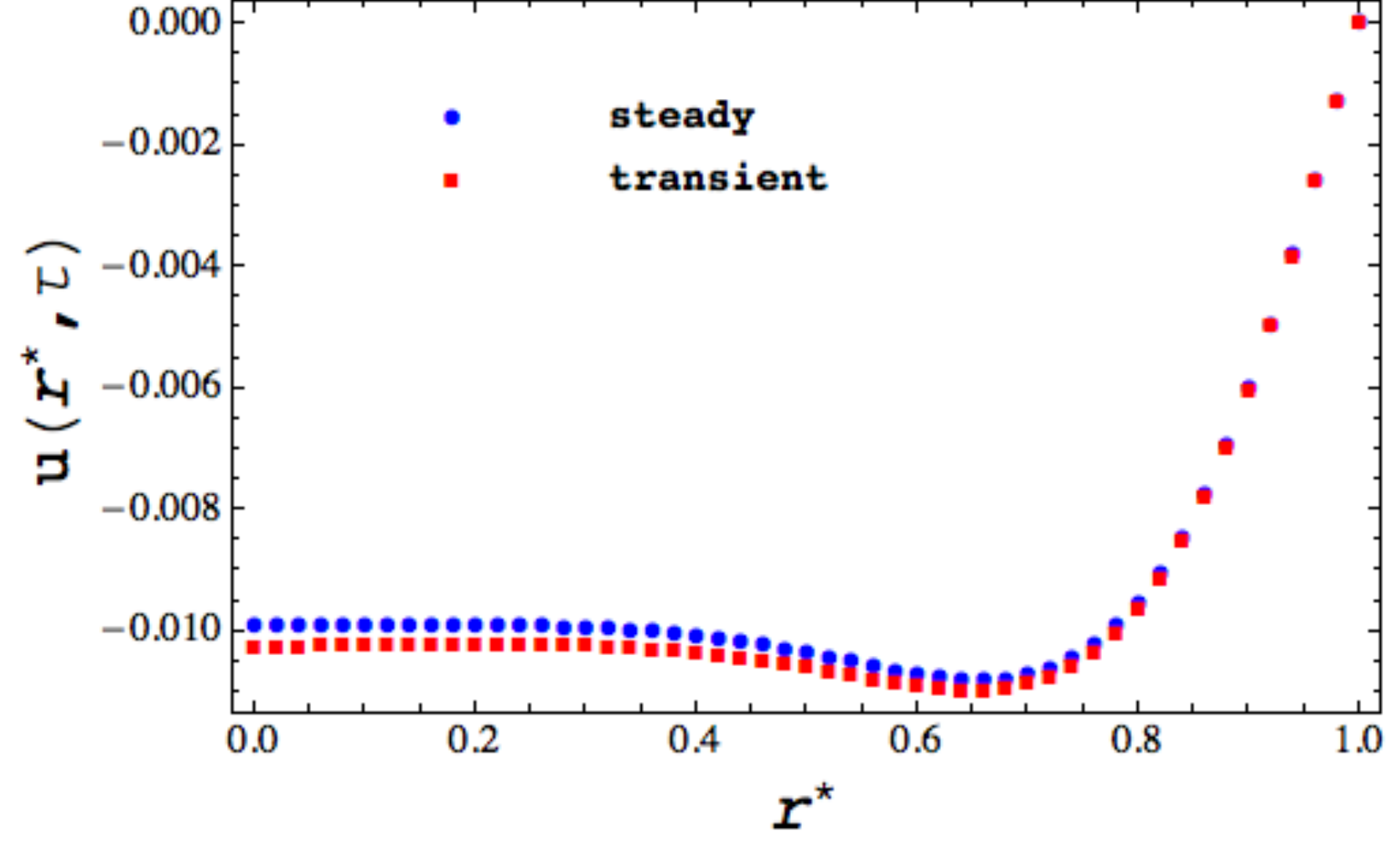}
\caption{Comparison of the Steady oscillating and transient solutions at the tipical time $\tau_1$  for a frequency $\omega=100$.}
	\label{fig:un}
\end{figure}

\section{Conclusions}

The precise numerical solution of the equation that defines the maximum of the overshoot from the mean squared velocity field showed that the location of the maximum in terms of $B_0$ changes with the kinematic reynolds number $\omega^*$ in the range of frequencies within the laminar regime. 

From the study of the Richardson's overshoot at different times it was identified the existence of some kind of transverse damped waves in circular pipe flow similar to those ones observed by Stokes in the problem of the flow generated by a oscillating infinite plate, this observation was used for the establishment of a semi-empirical law that gives the functional relation of the mean overshoot maximum with the kinematic reynolds number, say $B_0(\omega^*)$. 

A transient solution was constructed and verified asymptotically for large times, and  the tipical time for which the transient solution resembles the steady oscillating one was identified to be dependent of the viscosity of the fluid and of the radius of the pipe.

\end{document}